\documentclass[conference,10pt]{IEEEtran}

\long\def\beginpgfgraphicnamed#1#2\endpgfgraphicnamed{\includegraphics{#1}}

\usepackage{cite}
\usepackage[cmex10]{amsmath}
\usepackage{amssymb}
\usepackage{multirow}
\usepackage{graphicx}
\usepackage{mdwmath}
\usepackage{pgfplots}
\usepackage{array}	
\usepackage{calc}
\usepackage{xspace}
\usepackage{xcolor}
\usepackage{psfrag}
\usepackage{bm}
\usepackage{textcomp}
\usepackage{ifthen}
\usepackage{epsfig}
\usepackage{comment}
\usepackage{soul}
\usepackage{acronym}

\newsavebox\mybox

\pgfrealjobname{spawc2015}

\newcolumntype{C}[1]{>{\centering\arraybackslash$}p{#1}<{$}}

\newboolean{showfig}
\setboolean{showfig}{true}

\newboolean{showzero}
\setboolean{showzero}{false}

\newcommand{\includegraphicsLS}[2]{\ifthenelse{\boolean{showfig}}{{\includegraphics[#1]{#2}}}{\psdraft\includegraphics[#1]{#2}\psfull}}

\begin{document}

\renewcommand*{\acffont}[1]{\emph{#1}}
\acrodef{ADC}{Analog-to-Digital Converter}
\acrodef{BEC}{Binary Erasure Channel}
\acrodef{BER}{Bit Error Rate}
\acrodef{BTC}{Block Turbo Code}
\acrodef{DAC}{Digital-to-Analog Converter}
\acrodef{DMS}{Discrete Memoryless Symmetric}
\acrodef{DSP}{Digital Signal Processor}
\acrodef{DWDM}{Dense Wavelength Division Multiplex}
\acrodef{FEC}{Forward Error Correction}
\acrodef{FPGA}{Field-Programmable Gate Array}
\acrodef{LDPC}{Low-Density Parity-Check}
\acrodef{LLR}{Log-Likelihood Ratio}
\acrodef{MET}{Multi-Edge-Type}
\acrodef{ML}{Maximum Likelihood}
\acrodef{NCG}{Net Coding Gain}
\acrodef{ROADM}{Reconfigurable Add-Drop Multiplexer}
\acrodef{SC}{Spatially Coupled}

\makeatletter
\let\old@ps@headings\ps@headings
\let\old@ps@IEEEtitlepagestyle\ps@IEEEtitlepagestyle
\def\confheader#1{%
  \def\ps@headings{%
    \old@ps@headings%
    \def\@oddhead{\strut\hfill#1\hfill\strut}%
    \def\@evenhead{\strut\hfill#1\hfill\strut}%
  }%
  \def\ps@IEEEtitlepagestyle{%
    \old@ps@IEEEtitlepagestyle%
    \def\@oddhead{\strut\hfill#1\hfill\strut}%
    \def\@evenhead{\strut\hfill#1\hfill\strut}%
  }%
  \ps@headings%
}
\makeatother

\confheader{%
\scriptsize{Accepted for publication at the IEEE Workshop on Signal Processing Advances in Wireless Communications (SPAWC), June 2015, Stockholm, Sweden. \textcopyright IEEE}
}

%
\title{Spatially Coupled Codes and Optical Fiber Communications: An Ideal Match?}


\author{
\IEEEauthorblockN{Laurent Schmalen\IEEEauthorrefmark{1}, Detlef Suikat\IEEEauthorrefmark{1}, Detlef R\"osener\IEEEauthorrefmark{1}, Vahid Aref\IEEEauthorrefmark{1},  Andreas Leven\IEEEauthorrefmark{1}, Stephan ten Brink\IEEEauthorrefmark{2}}
\IEEEauthorblockA{\IEEEauthorrefmark{1}Alcatel-Lucent Bell Laboratories, Lorenzstr.  10, 70435 Stuttgart, Germany, {\tt first.last@alcatel-lucent.com}}
\IEEEauthorblockA{\IEEEauthorrefmark{2}University of Stuttgart, Institute of Telecommunications, Pfaffenwaldring 47, Stuttgart, Germany}
}

\markboth{IEEE Journal of Lightwave Technology, Vol. XX, No. YY, ZZZZZZ 2011}%
	 {Schmalen \emph{et. al}: Next Generation Error Correcting Codes for Lightwave Systems}

\maketitle
\begin{abstract}
 In this paper, we highlight the class of spatially coupled codes and discuss their applicability to long-haul and submarine optical communication systems. We first demonstrate how to optimize irregular spatially coupled LDPC codes for their use in optical communications with limited decoding hardware complexity and then present simulation results with an FPGA-based decoder where we show that very low error rates can be achieved and that conventional block-based LDPC codes can be outperformed. In the second part of the paper, we focus on the combination of spatially coupled LDPC codes with different demodulators and detectors, important for future systems with adaptive modulation and for varying channel characteristics. We demonstrate that SC codes can be employed as universal, channel-agnostic coding schemes.
\end{abstract}%

\begin{IEEEkeywords}
Error correction codes, Low-density parity-check codes, Spatial coupling, Optical Communications
\end{IEEEkeywords}

\section{Introduction}
\let\thefootnote\relax\footnotetext{Parts of this work were supported by the German Government in the frame of the CELTIC+/BMBF project SASER-SaveNet.}
Modern high-speed optical communication systems require high-performing \ac{FEC} implementations that support throughputs of 100 Gbit/s or multiples thereof, that have low power consumption, that realize  \acp{NCG} close to the theoretical limits at a target \ac{BER} of $10^{-15}$, and that are preferably adapted to the peculiarities of the optical channel. 

Especially with the advent of coherent transmission schemes and the utilization of high resolution \acp{ADC}, soft-decision decoding has become an attractive means of reliably increasing the transmission reach of lightwave systems. Currently, there are two popular classes of codes for soft-decision decoding that are attractive for implementation in optical receivers at decoding throughputs of 100 Gbit/s and above: \acf{LDPC} codes and \acp{BTC}. The latter can be decoded with a highly parallelizable, rapidly converging soft-decision decoding algorithm, usually have a large minimum distance, but require large block lengths of more than $100,000$\,bits to realize codes with small
overheads, leading to decoding latencies that can be detrimental in certain applications. With overheads of more than 15\% to 20\%, these
codes no longer perform well, at least under hard-decision decoding~\cite{Justesen}. \ac{LDPC} codes are understood and are suited to realize codes with lengths of a few $10,000$\,bits and overheads above $15$\%.

Recently, the class of \ac{SC} codes\cite{Kudekar0211} has gained widespread interest due to the fact that these codes are asymptotically capacity-achieving, have appealing encoding and decoding complexity and show outstanding practical decoding performance. \ac{SC} codes are an extension of existing coding schemes by a superimposed convolutional structure. The technique of spatial coupling can be applied to most existing codes, the most popular are however \ac{LDPC} codes~\cite{Kudekar0211} and \acp{BTC}~\cite{ZhangStaircase}, which have found use in optical communications (\emph{staircase codes}) and show outstanding performance, operating within  $0.5$\,dB of the capacity of the hard-decision AWGN channel. 

In this paper, we discuss the use of \ac{SC} codes in optical communications and especially focus on \ac{SC}-\ac{LDPC} codes. We summarize some recent advances and design guidelines for \ac{SC}-\ac{LDPC} codes and show by means of an \ac{FPGA}-based decoding platform that large gains at low bit error rates can be realized with relatively small codes when compared with state-of-the-art LDPC codes. The aim of this paper is to show that \ac{SC}-\ac{LDPC} codes are mature channel codes that are viable candidates for future optical communication systems with large \acp{NCG}. Furthermore, their universality makes them attractive for flexible transceivers with adaptive modulation.




\section{LDPC \& Spatially Coupled LDPC Codes}\label{sec:ldpc_sc}

An \ac{LDPC} code is defined by the null space of a  \emph{sparse} parity-check matrix $\bm{H}$ of size $\dim{\bm{H}} = M\times N$ where the code contains all binary code words $\bm{x}$ of length $N$ such that $\bm{H}\bm{x}^T=\bm{0}$, i.e., $\mathcal{C}_{\text{LDPC}} = \{\bm{x}\in\{0;1\}^N:\bm{H}\bm{x}^T=\bm{0}\}$. 

Each row of $\bm{H}$ is considered to be a \emph{check node}, while each column of $\bm{H}$ is usually termed \emph{variable node}. We say that the \emph{variable degree} (or \emph{variable node degree}) of a code is \emph{regular} with degree $d_v$ if the number of ``1''s in each column is constant and amounts to $d_v$. We say that the \emph{check degree} (or \emph{check node degree}) of a code is \emph{regular} with degree $d_c$ if the number of ``1''s in each row of $\bm{H}$ is constant and amounts to $d_c$. The class of \emph{irregular} \ac{LDPC} codes has the property that the number of ``1''s in each column and/or row is not constant.  The \emph{degree profile} of an irregular \ac{LDPC} code
indicates the fraction of columns/rows of a certain
degree. More precisely, $a_{v,i}$ represents the fraction of columns with
$i$ ``1''s (e.g., if $a_{v,3} = \frac{1}{2}$, half the columns of $\bm{H}$ have three ``1''s). Note that $\sum_i a_{v,i} = 1$ has
to hold. Similarly, $a_{c,i}$ represents the fraction of rows (i.e., checks) with $i$ ``1''s.

\ac{LDPC} codes form an important class of codes in optical communications~\cite{LevenSchmalen}. \ac{LDPC} codes with soft-decision decoding are currently being deployed in systems operating at 100\,Gbit/s and, e.g., utilizing 16 iterations~\cite{YamazakietalOptEx}.
Modern high-performance \ac{FEC} systems in optical communications
are sometimes constructed using a soft-decision \ac{LDPC}
inner code which reduces the \ac{BER} to a level of $10^{-3}$
to $10^{-6}$ and a hard-decision algebraic outer cleanup code which pushes
the system \ac{BER} to levels below $10^{-12}$~\cite{Miyata2009}. The outer cleanup code is used to combat the \emph{error floor} that is present in most \ac{LDPC} codes. Note that the implementation of a coding system with an outer cleanup code requires a thorough understanding of the \ac{LDPC} code and a properly designed interleaver between the \ac{LDPC} and the outer code. 
Recently, there has been some interest to avoid the use of an outer cleanup code and to use only soft-decision \ac{LDPC} codes with very low error floors, leading to coding schemes with less rate loss and less latency. With increasing computational resources, it is now also feasible to evaluate very low target \acp{BER} of \ac{LDPC} codes and optimize the codes to have very low error floors below the system's target \ac{BER}~\cite{ZhangGlobecom2006}. Although the internal data flow of an LDPC decoder may be larger by more than an order of magnitude~\cite{Smithxx12} than that of a BTC, several techniques can be used to lower the data-flow, e.g., the use of layered decoding~\cite{Hocevar2004} and min-sum decoding, requiring only two $q$-ary, $d_c+1$ binary and one $\lceil\log_2d_c\rceil$-ary message per check node. 



\ac{SC}-\ac{LDPC} codes were introduced more than a decade ago\cite{Felstroem0699}\footnote{Originally, these codes were called \emph{\ac{LDPC} convolutional codes}. The term ``spatially coupled'' has been introduced\cite{Kudekar0211} to denote the more general phenomenon of coupling several independent code(word)s, by a superimposed, convolutional-like structure.} but their outstanding properties have only been fully realized recently, when Lentmaier et al. noticed\cite{Lentmaier_T1} that the estimated decoding performance of a certain class of terminated protograph-based \ac{SC}-\ac{LDPC} codes with a simple message passing decoder is close to the performance of the underlying code ensemble under \ac{ML} decoding as $n$ grows, which was subsequently proven rigorously in \cite{Kudekar0211,Kudekar2012}, if certain particular conditions on the code structure are fulfilled.

A left-terminated \ac{SC}-\ac{LDPC} code is basically an \ac{LDPC} code with a structured, infinitely extended parity-check matrix
\begin{align}
{\small\bm{H}_{\text{SC}} = \left(\begin{array}{@{}ccccc@{}}
\bm{H}_0(0) &  &  &  &  \\[0.3em] 
\bm{H}_1(1) & \bm{H}_0(1) &  &  &  \\[-0.3em]
\vdots & \bm{H}_{1}(2) &  &  &  \\[-0.3em]
\bm{H}_\mu(\mu) & \vdots & \ddots &  &  \\[-0.7em] 
 & \bm{H}_{\mu}(\mu+1) & \ddots &  &  \\[-0.5em] 
 &  &  & \bm{H}_0(t) &  \\[-0.5em] 
 &  & \ddots & \bm{H}_1(t+1) & \ddots \\[-0.3em] 
 &  &  & \vdots & \ddots \\[-0.3em] 
 &  &  & \bm{H}_\mu(t+\mu) &  \\[-0.5em] 
 &  &  &  & \ddots
\end{array} \right)}\label{eq:timeinvarmatrix}
\end{align}
with $\bm{H}_i(t)$ being sparse binary parity-check matrices with $\dim\bm{H}_i(t)= m\times n$  and $\mu$ denoting the \emph{syndrome former memory} of the code. Every code word $\bm{x}$ of the code has to fulfill $\bm{H}_{\text{SC}}\bm{x}^T=\bm{0}$. One advantage of \ac{SC}-\ac{LDPC} codes is that the infinitely long code words can conveniently be decoded with acceptable latency using a simple windowed decoder~\cite{Iyengar12}.  In practice, in order to construct codes of finite length, e.g., to adhere to certain framing structures in the communication system at hand, the infinitely extended matrix $\bm{H}_{\text{SC}}$ is \emph{terminated} resulting in finite length code. One example of termination is \emph{zero-termination}, where the matrix $\bm{H}_{\text{SC}}$ is cut off after $t=L$ parts, resulting in a code of length $N=Ln$ and a parity-check matrix $\bm{H}_{\text{SC}}^{[\text{Term.}]}$ of size $\dim\bm{H}_{\text{SC}}^{[\text{Term.}]} = (L+\mu)m\times Ln$. Note that this termination leads to a \emph{rate loss}, which can however be kept small if $L$ is chosen large enough. For a discussion of termination schemes, we refer the interested reader to~\cite{LevenSchmalen,HaegerECOC14}.


\ac{SC} codes are now emerging in various applications. Two examples of \ac{SC} product codes are the staircase code\cite{Smithxx12} and the braided BCH codes~\cite{Jian1213}, 
for hard-decision decoding in optical communications. SC-LDPC codes
may also be viable for pragmatic coded modulation schemes~\cite{Schmalen2013,HaegerECOC14}.

In order to simplify the design of hardware, we first drop the time dependency and only consider the time-independent (left-terminated) parity-check matrix with $\bm{H}_i(t) = \bm{H}_i$, $\forall i\in\{0,1,\ldots, \mu\}$,
which is attractive for implementation as the sub-matrices $\bm{H}_i$ can be easily reused in the encoder and decoder hardware.
In this time-invariant construction with $\dim\bm{H}_i = m\times n$, we can give the following upper bound on the minimum distance of the code~\cite[Eq.~(7)]{LevenSchmalen}
\begin{align}
d_{\min} \leq (m+1)(\mu m + 1)\,.\label{eq:distbound}
\end{align}
To construct codes with large enough minimum distances, 
we maximize the size of the sub-matrices $\bm{H}_i$, i.e., $m$, which has a quadratic influence on~\eqref{eq:distbound}. In order to keep the complexity of the so-constructed code small, we restrict ourselves to small values of the syndrome former memory $\mu$ with either $\mu=1$ or $\mu=2$.  We call such codes \emph{weakly coupled} codes~\cite{SchmalenJLTSC15}.

\section{Rapidly Converging \ac{SC}-\ac{LDPC} Codes}\label{sec:weakly}

In the past, irregular block \ac{LDPC} codes have been used to design codes that perform very well for low SNRs, but these schemes do sometimes suffer from relatively high error floors requiring the use of an outer code that leads to inherent rate losses.  In the case of \ac{SC}-\ac{LDPC} codes, we can use the irregularity to control the propagation speed of the decoding wave of a windowed decoder, i.e., we can minimize the number of iterations $I_{\text{req}}$ that are necessary until a windowed decoder can advance by one step~\cite{Aref1013}. To simplify the code construction and to illustrate the concept, we only use the most simple form of irregularity and construct slightly irregular \ac{SC}-\ac{LDPC} codes with \mbox{degree-3} and additionally with either degree-4 or {degree-6} variable nodes. We avoid degree-2 variables nodes due to their potentially detrimental effect on the error floor. Also, in contrast to block LDPC codes, degree-2 nodes are \emph{not} of the same importance for \ac{SC}-\ac{LDPC} codes. We vary the fraction of degree-4 or degree-6 nodes between 0 and 1 and select the check nodes such that a rate $r=\frac{4}{5}$ (25\% overhead) code is constructed. 
We perform full density evolution using the irregular version of Kudekar's $(d_v, d_c, w, L)$ ensemble~\cite{Kudekar0211} for random spatial coupling with $w=3$ using an AWGN channel and measure the required $E_{\text{b}}/N_0$ values to advance the decoding wave by $I_{\text{req}}$ steps.

\begin{figure}[tp]
   \centering
   \beginpgfgraphicnamed{fig_1}
   \input{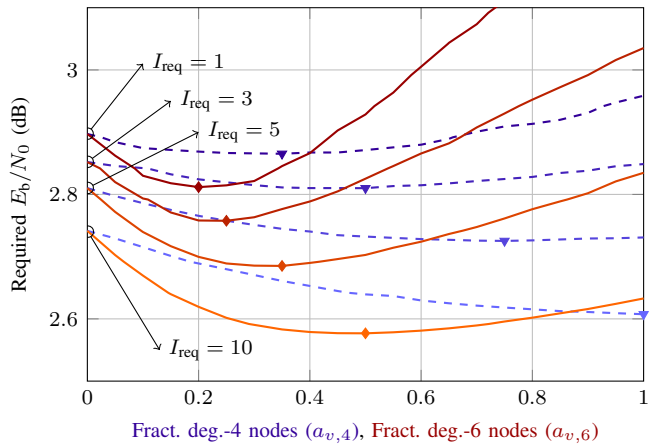}
   \endpgfgraphicnamed
    \caption{Required $E_{\text{b}}/N_0$ to operate a windowed decoder with $I_{\text{req}}$ iterations per segment for slightly irregular weakly coupled \ac{SC} codes with degree-3 nodes and additionally degree-4 (dashed lines) or degree-6 (solid lines) variable nodes. (results obtained by density evolution)}
    \label{fig:speed}
\end{figure}

The density evolution results are shown in Fig.~\ref{fig:speed} for varying $a_{v,4}$ and $a_{v,6}$. We can see that using additionally degree-4 (besides degree-3) variables does not lead to noteworthy gains, which is why we focus on additional degree-6 nodes in this paper. The convergence speed improves by selecting a proper value of $a_{v,6}$ leading to a smaller required $E_{\text{b}}/N_0$. The selection depends however on $I_{\text{req}}$. As we intend to construct low complexity decoders with $I_{\text{req}}\in \{1,3\}$, we can see that in this case, the optimum is achieved with  $a_{v,6}=0.2$ (20\% of degree-6 variable nodes, 80\% of degree-3 variable nodes).
We can see that by proper selection of $a_{v,6}$, we can obtain codes that have an improved decoding convergence, however, we also see that depending on the selection, a worse convergence behavior than for the regular case can result.
We also observe that if we want a code that operates extremely close to capacity, the optimum value of $a_{v,6}$ is larger (around $0.5$) than for the more practical case, where the optimum lies at $a_{v,6}=0.2$.
Note that although we use Kudekar's ensemble for density evolution, the codes we construct in the next section are generated from protographs, similar to those in~\cite{Lentmaier_T1}, as these exhibit better finite length performance.

\newcommand{\sfx}{\ensuremath{\mathsf{x}}\xspace}
\newcommand{\sfy}{\ensuremath{\mathsf{y}}\xspace}

\subsection{FPGA-based Verification}\label{sec:weakly_fpga}
\begin{figure}[t!]
\includegraphics[width=\columnwidth]{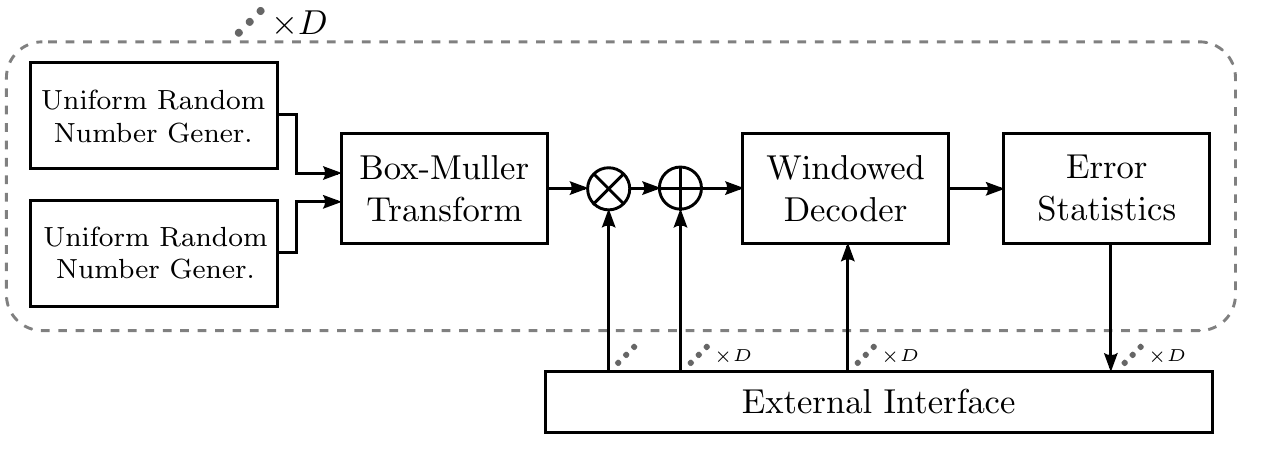}
\caption{High-level schematic of the \ac{FPGA} evaluation platform.}
\label{fig:fpga_schematic}
\end{figure}

In order to verify the performance of the rapidly converging weakly coupled \ac{SC}-\ac{LDPC} codes, we use a \acf{FPGA} platform, whose high-level diagram is illustrated in Fig.~\ref{fig:fpga_schematic}~\cite{SchmalenJLTSC15}. This platform is similar to other platforms reported in the literature~\cite{ZhangGlobecom2006} and consists of three parts: A Gaussian noise generator, an \ac{FEC} decoder and an error detecting circuit. The Gaussian noise generator generates Gaussian distributed \acp{LLR}, stemming from BPSK transmission over an AWGN channel, using uniform random number generators and the Box-Muller transform. These are then fed to the \ac{LDPC} decoder after quantization to 15 levels. The \ac{LDPC} decoder is based on the layered decoding algorithm~\cite{Hocevar2004} and uses a scaled-minsum check node computation rule with constant scaling factor. 

The windowed decoder that is implemented can be sub-divided into three steps. In the first step, a new sub-block of $n$ quantized \acp{LLR} is received from the random number generator and put into the vacant position of the decoder's \ac{LLR} memory. Decoding takes place by considering $W=13$ copies of $\left(\bm{H}_{\mu}\ \bm{H}_{\mu-1}\ \cdots\ \bm{H}_0\right)$. The windowed decoded considers an equivalent matrix
of size $Wm\times (W+\mu-1)n$ which it processes before shifting in $n$ new values. In order to maximize the hardware utilization, within a window, we use two parallel decoders that operate on non-overlapping portions of that matrix. In a first step, the first decoding engine operates on the first $m$ check nodes of the matrix under consideration while the second engine operates in parallel on the $m$ check nodes starting at position $6m$. In general, the first engine processes the check nodes at position $i\in[1,Wm]$ while the second engine processes the check node $(i+6m-1)\mod Wm + 1$.  Note that only a single iteration is carried out to guarantee the required throughput, corresponding effectively to $2W$ iterations per bit (due to the use of two engines).

\begin{figure*}
\begin{align}
x_{i}^{(\ell)} = \left(1 -f_D\left(1 - L \left(1-\frac{1}{w}\sum_{j=0}^{w-1}\left(1-\frac{1}{w}\sum_{k=0}^{w-1}x_{i-j+k}^{(\ell-1)}\right)^{d_c-1}\right) \right) \right)\cdot \lambda\left(1-\frac{1}{w}\sum_{j=0}^{w-1}\left(1-\frac{1}{w}\sum_{k=0}^{w-1}x_{i-j+k}^{(\ell-1)}\right)^{d_c-1}\right)\label{eq:de_kudekar}
\end{align}
\hrule
\end{figure*}

\begin{figure}[t!]
\beginpgfgraphicnamed{fig_2}
\input{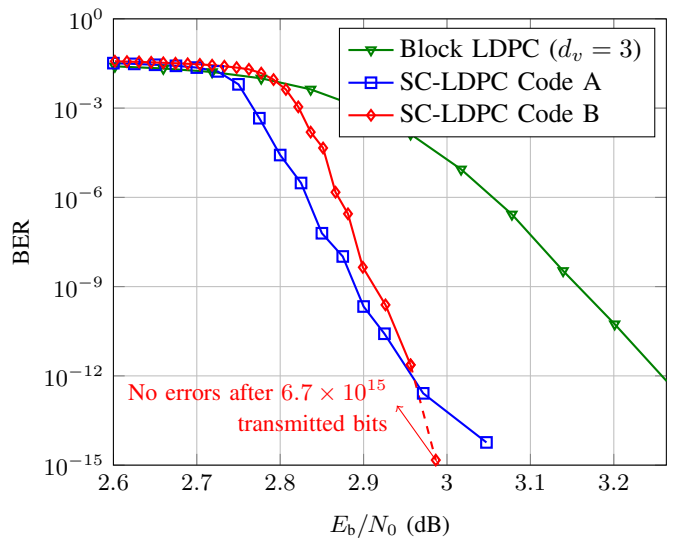}
\endpgfgraphicnamed

\caption{Simulation results with \ac{FPGA}-based windowed decoding, $W=13$, two decoder instances.}
\label{fig:fpga_simres}
\end{figure}

The output of the \ac{LDPC} decoder is connected to the \ac{BER} evaluation unit, which counts the bit errors and reports the error positions. We use Virtex-7 \acp{FPGA} allowing for a throughput of several Gbit/s to evaluate the BER performance of several coding schemes of rate $r=\frac{4}{5}=0.8$, i.e., of 25\% coding overhead. We select this particular rate due to its importance in today's~\ac{DWDM} systems. Current and future $100$\,Gbit/s (with QPSK) or $200$\,Gbit/s (with 16-QAM) systems are often operated in $50$\,GHz channels with an exploitable bandwidth of roughly $37.5$\,GHz due to \acp{ROADM} with non-flat frequency characteristic. With almost rectangular pulse shapes (root-raised cosine with small roll-off $\alpha$) and today's generation of \acp{DAC}, symbol rates of $32$\,GBaud can be realized. With dual-polarization QPSK transmission, gross bit rates of 128\,Gbit/s can be realized. Assuming signaling and protocol overheads of 3\,Gbit/s, this leads to a code that adds 25\,Gbit/s parity overhead (i.e., of rate $r=0.8$). We compare three codes: 
\begin{itemize}
\item[$\nabla$]	As reference, we consider a regular block QC-LDPC code (marker $\nabla$) with variable node degree $d_v=3$ and check node degree $d_c=15$. The code is a quasi-cyclic code of girth 10 and block length $N=31,200$, constructed using cyclically shifted identity matrices of size $32\times 32$ and decoded with $I=26$ row-layered iterations.
\item[$\square$] \ac{SC}-\ac{LDPC} Code A ($\square$) is the rapidly converging irregular code with syndrome former memory $\mu=2$, $a_{v,6}=0.2$ and $a_{v,3}=0.8$ and check node degree $d_c=18$. The sub-block size is $n=7500$ ($\dim\bm{H}_i = 1500\times 7500$). 
\item[$\lozenge$] \ac{SC}-\ac{LDPC} Code B ($\lozenge$) is a regular $d_v=4$ code with $d_c=20$ and syndrome former memory $\mu=1$. The size of the sub-matrices is identical to those of \ac{SC}-\ac{LDPC} code A, however, we select $\mu=1$.
\end{itemize}
Both \ac{SC} codes are constructed from cyclic permutation matrices of size $30\times 30$ and are terminated after $L=90$ subblocks.  The simulation results are shown in Fig.~\ref{fig:fpga_simres}. The block code, which has a matrix that has been optimized for low error floors, is outperformed by both \ac{SC}-\ac{LDPC} codes.
\ac{SC}-\ac{LDPC} code A offers a coding gain of around 0.3\,dB at a BER of $10^{-12}$ compared to the conventional block LDPC code, but an error floor starts to manifest. This error floor is not due to any trapping sets, but due to a few uncorrected bits after windowed decoding, which can be recovered with a few-error correcting outer code. Code B has a BER curve that starts to decay at worse channels, but the BER curves cross at $\approx 10^{-13}$. For the next simulated point, we did not observe any bit errors, and hence we conjecture a lower error error floor than for Code A. Note that no special measures have been taken to combat an error floor: only a plain scaled min-sum decoder has been used. With the block code, post-processing~\cite{ZhangGlobecom2006} may be necessary to combat the error floor. 

Another advantage of \ac{SC}-\ac{LDPC} codes is that they are future-proof: While the block code does not benefit from further decoding iterations, as its performance is already close to its decoding threshold, the scaling behavior of the SC-LDPC code allows to carry out further iterations and achieve still larger coding gains, as the gap to the decoding threshold is still non-negligible. This makes these codes attractive for standardization.

\definecolor{mycolor3}{rgb}{0.20000,0.00000,0.60000}%
\definecolor{mycolor5}{rgb}{0.73333,0.13333,0.00000}%

\section{SC-LDPC Codes for Modulation and Detection}

As future optical networks tend to become increasingly flexible and \emph{elastic}, 
transceivers that integrate a certain amount of flexibility with respect to coding and modulation formats are required. Especially the modulation format is expected to change when transceivers are designed for long-haul or short-haul applications, where the latter require high spectral efficiencies (e.g., data center interconnects). In this section, we show
that \ac{SC} codes are perfectly suited to be combined with varying modulation formats due to their \emph{universality} properties~\cite{Kudekar2012}. We combine~\ac{SC}-\ac{LDPC} codes with a modulator and use \emph{density evolution} to show how the detector front-end influences the performance of the codes. In conventional (block) \ac{LDPC} code design, usually
the code needs to be ``matched'' to the transfer curve of the
detection front-end~\cite{tenBrink2004}. If the code is not well matched to the front-end,
a performance loss occurs. If the detector front-end has highly
varying characteristics, due to, e.g., varying modulation formats or channels, several codes would need to be implemented and selected depending on the conditions, which is not feasible  in optical networks, where feedback is usually difficult to realize and where different codes cannot be implemented due to hardware constraints.

In contrast to many block LDPC codes, spatially coupled LDPC codes can converge below the pinch-off in the EXIT chart due to the effect of threshold saturation~\cite{Kudekar0211}. Hence, even if the code is not well matched to the demodulator/detector from a traditional point of view, we can hope to successfully decode. We can hence use a single code which is \emph{universally} good in all scenarios and the code design can stay \emph{agnostic} to the channel/detector behavior. In order to illustrate the concept, we model the detector by a linear EXIT characteristic
\[
I_E^{[D]} = f_D(I_A^{[D]}) = a\cdot I_A^{[D]} + I_C - \frac{a}{2}
\]
where $a$ controls the slope of the characteristic and $I_C = \int_0^1f_D(\tau)d\tau$ describes the mutual information of the communication channel. The slope $a$ models the effect of e.g., different modulation formats, different bit labelings in higher order modulation and different detectors. We assume that the output of the detector can be modeled using a \ac{BEC}. There therefore also use \ac{BEC} message passing. We compare two different code approaches; first we use the spatially coupled $(d_v,d_c,w,L)$ ensemble presented in~\cite{Kudekar0211} with the density evolution equation for iterative detection given by~\eqref{eq:de_kudekar} where $L(\xi) = \sum_i a_{v,i}\xi^i$ denotes the node-perspective degree distribution polynomial, $\lambda(\xi) = L^\prime(\xi)/L^\prime(1)$ the edge-perspective degree distribution, and $x_i^{(\ell)}$ the edge message erasure probability of spatial position $i$ at iteration $\ell$. Additionally, we generate protograph based codes end employ \ac{MET} density evolution~\cite{Lentmaier_T1} including iterative detection. We consider two code families of rate $0.8$: The first family is the rapidly converging code from Sec.~\ref{sec:weakly} with $L(x) = \frac{4}{5}x^3+\frac{1}{5}x^6$ and $d_c=18$ where we use $w=3$ and $L=100$ in Kudekar's ensemble and $\bm{B}_0 = \bm{B}_1 = \bm{B}_2 = \begin{pmatrix} 2 & 1 & 1 & 1 & 1 \end{pmatrix}$ with $\mu=3$ in the protograph ensemble. The second code is a regular code where we use Kudekar's $1(d_v=4,d_c=20,w=3,L=100)$ ensemble and a protograph ensemble with $\bm{B}_0=\begin{pmatrix} 1 & 2 & 1 & 2 & 1\end{pmatrix}$ and $\bm{B}_0=\begin{pmatrix} 3 & 2 & 3 & 2 & 3\end{pmatrix}$ with $\mu=2$~\cite{Schmalen2013}.

\newcommand\solidrule[1][0.7cm]{\rule[0.5ex]{#1}{.4pt}}
\newcommand\dashedrule{\mbox{%
  \solidrule[1mm]\hspace{1mm}\solidrule[1mm]\hspace{1mm}\solidrule[1mm]}}

\begin{figure}[t!]
\beginpgfgraphicnamed{fig_3}
\input{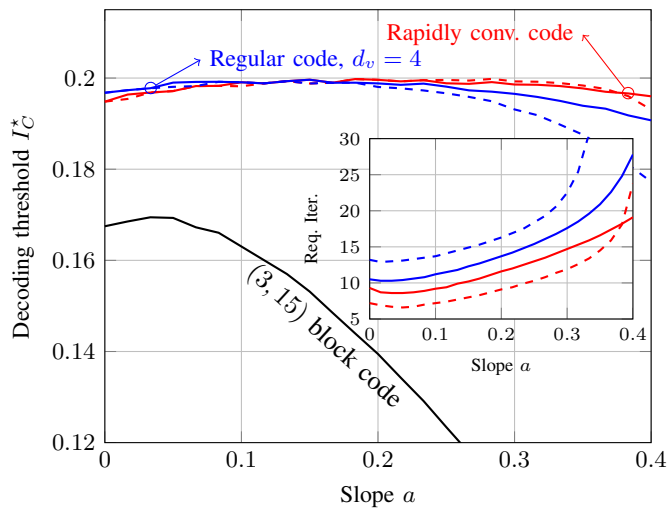}
\endpgfgraphicnamed
\caption{Decoding thresholds of different SC-LDPC codes for varying detector characteristics with varying slope $a$ and a regular $(3,6)$ block code.}
\label{fig:det_thresholds}
\end{figure}

Figure~\ref{fig:det_thresholds} shows the DE results where we use solid lines (\solidrule) to show the decoding thresholds for Kudekar's ensemble and dashed lines (\protect\dashedrule) for the protograph-based ensemble. All SC codes have decoding thresholds close to the theoretical limit of $I_{C,\max} = 0.2$ and the decoding threshold is almost independent of the detector characteristic's slope $a$. A regular block LDPC code has a highly varying threshold for different slopes $a$. The flat threshold behavior for \ac{SC}{LDPC} codes indicates a \emph{universal}, \emph{channel-agnostic} behavior. Even an optimized irregular LDPC code will only be good for a single slope parameter~\cite{PfluegerSCC13}. In order to improve the decoding threshold, we may deliberately select a precoder that has an EXIT characteristic with slope $a>0$, however, as the inset of Fig.~\ref{fig:det_thresholds} shows, the slope affects the decoding speed (measured at $I_C=0.185$), i.e., the number of iterations required to advance the decoding wave by one step, so that the complexity will grow alongside. For the case of the rapidly converging code,  $a>0$ further increases the decoding speed.

We have presented an example of such a system with differential detection ($a\approx 0.2$) that is adapted to a channel with varying phase noise in~\cite{SchmalenOFC15}. Therein, a single spatially coupled code was able to outperform two different LDPC codes optimized for different channel characteristics.

\section{Conclusions}

In this paper, we have highlighted \acf{SC}-LDPC codes as potential candidates for future lightwave transmission systems. We have optimized \ac{SC}-\ac{LDPC} codes for convergence speed and shown by means of an FPGA-based simulation that very low error rates can be obtained. Finally, we have shown that \ac{SC}-\ac{LDPC} can be good candidates if they employed in a system with iterative decoding and detection: a single code can be used in various channel conditions.




\end{document}